\documentclass[sigconf]{acmart}
\acmConference[ICCAD'26]{International
  Conference on Computer-Aided Design}
  {October 2026}{San Francisco, CA, USA}

\renewcommand\footnotetextcopyrightpermission[1]{}

\usepackage{algorithm}
\usepackage{graphicx}
\usepackage{enumitem}
\usepackage{indentfirst}
\usepackage{multirow}
\usepackage{subfigure}
\usepackage{amsmath,amsfonts}
\usepackage{algorithmic}
\usepackage{textcomp}
\usepackage{xcolor}
\usepackage{booktabs}
\usepackage{subcaption}
\usepackage{caption}
\usepackage{booktabs}
\usepackage{multirow}
\usepackage{makecell}
\usepackage{adjustbox}
\newcommand{\tblfont}{\fontsize{7.1}{8.2}\selectfont}

\begin{document}

\title{AutoDRI: Bridging the Semantic Gap for Automated Design Rule Integration in CP-SAT-Based Cell Synthesis under Multi-Patterning}


\author{Zihao Chen}
\affiliation{%
  \institution{Fudan University}
  \city{Shanghai}
  \country{China}}
\email{zihaochen17@m.fudan.edu.cn}

\author{Chung-Kuan Cheng}
\affiliation{%
  \institution{University of California San Diego}
  \city{San Diego}
  \state{California}
  \country{USA}}
\email{ckcheng@ucsd.edu}

\author{Yuhao Ren}
\affiliation{%
  \institution{Fudan University}
  \city{Shanghai}
  \country{China}}
\email{yhren24@m.fudan.edu.cn}

\author{Yucheng Wang}
\affiliation{%
  \institution{University of California San Diego}
  \city{San Diego}
  \state{California}
  \country{USA}}
\email{yuw132@ucsd.edu}

\author{Zhiang Wang}
\affiliation{%
  \institution{Fudan University}
  \city{Shanghai}
  \country{China}}
\email{zhiangwang@fudan.edu.cn}

\renewcommand{\shortauthors}{Ren et al.}


\begin{abstract}
Design-rule integration (DRI) remains a major bottleneck for scalable (Constraint Programming with SAT) CP-SAT-based standard cell synthesis and rapid technology enablement at advanced nodes. 
It still depends heavily on manual effort and domain expertise. 
Moreover, existing low-level rule encodings are not expressive enough for emerging constraints such as cut-based rules under multi-patterning technology.
This paper presents \textbf{AutoDRI}, a multi-agent framework for automated design-rule integration in standard cell synthesis. 
AutoDRI combines a geometric semantic library, a standardized conflict-set encoding, a constructive multicolor-cut modeling method, and a feedback-driven multi-agent flow to bridge the semantic gap between natural-language design rules and executable CP-SAT constraints. 
In the reported experiments, AutoDRI achieves near-perfect rule-integration correctness across 41 cell benchmarks under 10+ complex rules, including colored cut-mask spacing rules, reaching 33/33 correct integrations with Gemini-3-pro and 32/33 with GPT-5.4, while maintaining runtime comparable to manual hard-coding and passing KLayout DRC and Cadence LVS.
\end{abstract}


\maketitle

\section{Introduction}
\label{sec:intro}

\begin{figure}[htbp]
  \centering
  \includegraphics[width=1.03\columnwidth]{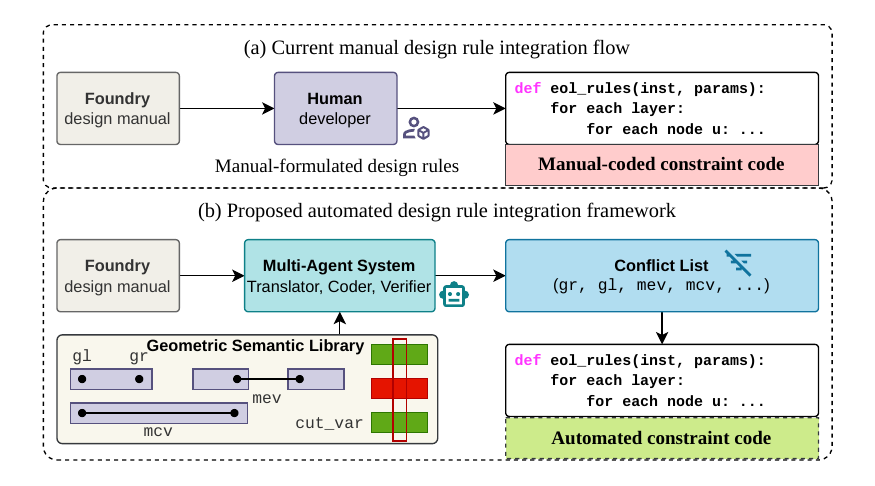} 
  \caption{
  Comparison of manual and AutoDRI rule-integration flows.}
  \label{fig:flow_comparation}
  \vspace{-8pt} 
\end{figure}

Automated standard cell synthesis can significantly accelerate the turnaround time for Design-Technology Co-optimization (DTCO) and System-Technology Co-optimization (STCO) while approaching optimal Power-Performance-Area (PPA) trade-offs~\cite{jacob_scaling_2017,kim_imec_2018}. 
Constraint-Optimization (CO)-based synthesis~\cite{CPCellUCSD,lee_spr_2021,van_cleeff_bonncell_2020,park_road_2019,li_nctucell_2019,cheng_gear-ratio-aware_2024} is particularly appealing: once design rules are correctly encoded into a simultaneous placement-and-routing formulation, every feasible solution is DRC-clean by construction~\cite{cheng_standard_2025}. However, this paradigm hinges on a largely manual step—\textbf{the translation of foundry design rules into executable solver constraints}. 
A key limitation of existing layout synthesis tools is their tight coupling to a single PDK, making cross-node migration labor-intensive. 
We identify three consequences of this rigidity. 
First, as foundries introduce new architectural features, collaboration with tool developers becomes necessary, yet critical process parameters are often obfuscated, forcing developers to reverse-engineer design intent. 
Second, a persistent temporal gap separates industrial practice from academic tooling; by the time academic frameworks adapt to a new technology node, industry has already completed DTCO manually, leaving academic tools outdated. 
Third, rule encoding requires substantial expertise and lacks a standardized protocol, making it difficult to transfer across tools.
Together, these factors make design-rule integration (DRI) the
primary bottleneck for rapid technology enablement under DTCO. 
The human-in-the-loop rule encoding process must be repeated,
debugged, and re-validated whenever a new node or rule revision
arrives, as illustrated by the current manual flow in
Figure~\ref{fig:flow_comparation}.

Recent advances in Large Language Models (LLMs) offer a natural path toward automating this bottleneck. LLMs have demonstrated strong reasoning and code-generation capabilities~\cite{roziere_code_2024, wei_chain--thought_nodate}, and a growing body of work explores LLM-based agents for complex engineering workflows~\cite{yao_evoplace_2026, wang_voyager_2023}. In the EDA domain, DRC-Coder~\cite{chang_drc-coder_2025} uses autonomous agents with Retrieval-Augmented Generation to produce commercial-grade DRC checker code, while D2D-LLM~\cite{tang_d2d-llm_2025} studies bidirectional translation between rule manuals and DRC decks. However, these efforts focus on \emph{procedural} artifacts for post-design physical verification—scripts that flag violations after layout generation. 
Rule integration for \emph{constructive} CP-SAT-based synthesis is fundamentally different: it requires translating natural-language design rules into declarative pseudo-Boolean constraints that remain logically consistent with a formal optimization model. Directly generating such constraints over raw grid coordinates leads to hallucination and logical mismatch, as the model must simultaneously reason about spatial geometry, Boolean logic, and solver-specific encoding conventions without a structured intermediate representation. We refer to this disconnect—between high-level rule semantics and low-level solver variables—as the \emph{semantic gap}.


The semantic gap has two facets. 
First, \textbf{there is no standardized rule-representation paradigm}. 
Existing formulations tightly couple rule semantics to grid-level coordinates, with no reusable abstraction layer between geometric intent and low-level encoding. 
As a result, each rule must be hand-derived and hard-coded, making the process opaque to both LLM agents and new human developers.

Second, \textbf{existing variables have limited expressiveness}. 
They are mainly tailored to minimum area and end-of-line distance. 
Commercial sub-7nm nodes increasingly rely on cut-based multi-patterning~\cite{chava_standard_2015}, where rules are defined over physical cut-mask shapes whose color assignment and spatial extent go beyond what existing primitives can capture~\cite{colorcut}.

This paper presents \textbf{AutoDRI}, a multi-agent framework
(Figure~\ref{fig:flow_comparation}) that closes both the
representation gap and the automation gap for design-rule integration
in CP-SAT-based standard cell synthesis.
AutoDRI introduces a layered abstraction stack---from
geometric semantics through conflict-set normalization to LLM-driven
code generation---so that adding or updating a design rule requires
only its natural-language specification rather than solver-level
expertise. 
Our contributions are as follows:

\begin{itemize}[noitemsep, topsep=0pt, leftmargin=*]
\item A geometric semantic library of reusable interval-level variables
that decouple rule semantics from grid-level coordinates, simplifying
the encoding of multi-conditional constraints and enhancing model
extensibility.

\item A constructive multicolor-cut modeling method that 
builds physical cut-mask representations from primitive semantic
variables, enabling CP-SAT-based synthesis under multi-patterning
technologies without modifying the underlying solver formulation.

\item A standardized conflict-set encoding that represents every design
rule as a normalized set of local forbidden Boolean patterns, providing a
structured template for systematic translation from natural-language
descriptions to executable constraints.

\item A multi-agent automated integration flow with Teacher--Translator Loop and Verifier Feedback that translates natural-language design
rules into executable CP-SAT code with minimal human intervention.

\item Validation on 41 cell benchmarks under 10+ complex rules,
demonstrating a nearly 100\% pass rate in KLayout DRC and Cadence LVS with
solving efficiency comparable to manual hard-coding.
\end{itemize}

The rest of this paper is organized as follows. Section~\ref{sec:prelim} provides the necessary preliminaries and reviews related work. Section~\ref{sec:semantic_representation} presents the semantic rule representation. Section~\ref{sec:automated_rule_integration} presents our automated design rule integration flow. Section~\ref{sec:experiment} discusses the experimental results. Finally, we conclude the paper in Section~\ref{sec:conclusions}.

\section{Preliminaries}
\label{sec:prelim}

This section defines the modeling abstractions used throughout the paper.
Given a set of natural-language design rules and their associated parameters, our framework translates them into executable solver constraints and integrates them into the CPCell framework~\cite{CPCellUCSD}. 
The output is a synthesized standard cell layout that satisfies the entire rule set simultaneously. 

We first describe the routing-grid graph structure assumed by the synthesis engine, then classify design rules in current commercial PDKs and discuss the limitations of existing rule representations.
Table~\ref{tab:notations} summarizes the key symbols used in this paper.

\begin{table}[hbtp]
\centering
\caption{Terminology and notation.}
\label{tab:notations}
\vspace{-0.4em}
\begin{tabular}{p{0.26\columnwidth} p{0.62\columnwidth}}
\toprule
\textbf{Notation} & \textbf{Meaning} \\
\midrule
$G=(V,E)$ & 3D routed grid graph \\
$u=(\ell,r,c)$ & Node on layer $\ell$, row $r$, column $c$ \\
$P(u,v)$ & Set of same-track edges between nodes $u$ and $v$ \\
$\chi(u)$ & Intrinsic color of node $u$ \\
$\kappa$ & Color label \\
$\texttt{gl}(u), \texttt{gr}(u)$ & Left/right boundary markers \\
$\texttt{gf}(u), \texttt{gb}(u)$ & Front/back boundary markers \\
$\texttt{mcv}(u,v)$ & Metal continuation variable \\
$\texttt{mev}(u,v)$ & Metal empty variable \\
$\texttt{cut\_var}(u)$ & Color-independent cut-selection variable \\
$c^{\mathrm{cut}}_\kappa(u)$ & Color-$\kappa$ cut-mask variable \\
$h(u_1,u_2)$ & Metal-breaking interval \\
$\mathcal{Q}(u_1,u_2)$ & Candidate cut-node set in a metal-breaking interval \\
$\mathcal{N}_\kappa(u)$ & Neighboring color-$\kappa$ nodes supporting cut extension \\
$\Phi_\rho$ & conflict set of design rule $\rho$ \\
$\boldsymbol{\gamma}_j$ & One local forbidden Boolean pattern \\
\bottomrule
\end{tabular}
\vspace{-0.8em}
\end{table}

\subsection{Routing-Grid Graph Structure}

We model the cell layout on a three-dimensional routing grid
$G = (V, E)$.  Each node $u \in V$ is a tuple $(\ell, r, c)$
denoting a grid point on metal layer~$\ell$ at row~$r$ and
column~$c$.  Routing is unidirectional: M0 and M2 carry horizontal
tracks while M1 carries vertical tracks.  The edge set~$E$ contains
an \emph{intra-layer} edge between every pair of adjacent nodes
along the routing direction of their shared layer, and an
\emph{inter-layer} edge (via) between every pair of vertically
adjacent nodes on neighboring layers.  A Boolean variable on each
edge indicates whether that edge is occupied by a metal segment or
a via. 
For two nodes $u, v$ on the same track, we write
$P(u,v) \subseteq E$ for the set of consecutive intra-layer edges
connecting $u$ to~$v$ along that track.  All geometric semantic
variables and design-rule constraints defined in subsequent sections
are expressed over~$V$, $E$, and the derived paths~$P$. 
This graph structure is shared by prior CO-based synthesis
frameworks such as SP\&R~\cite{lee_spr_2021} and
CPCell~\cite{CPCellUCSD}.

\begin{figure}[b]
  \centering
  \includegraphics[width=0.50\textwidth]{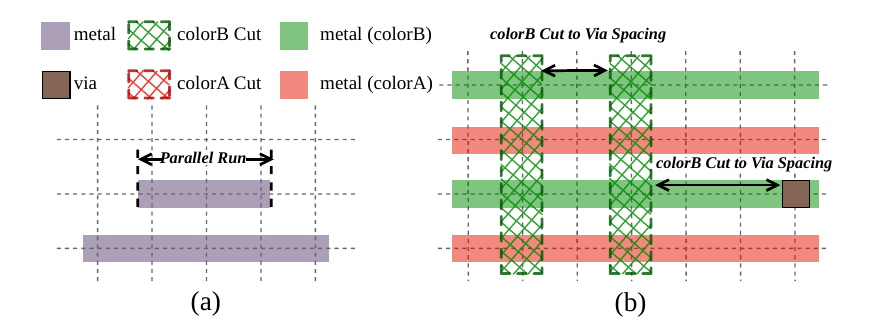}
  \caption{
    Representation limitations of existing encodings: (a) incomplete coverage of violation scenarios in context-dependent metal-length rules, and (b) inability to represent colored cut-mask spacing rules.
  }
  \label{fig:representation_limitation}
  \vspace{-0.5em}
\end{figure}

\begin{figure*}[htbp]
  \centering
  \includegraphics[width=1.0\textwidth]{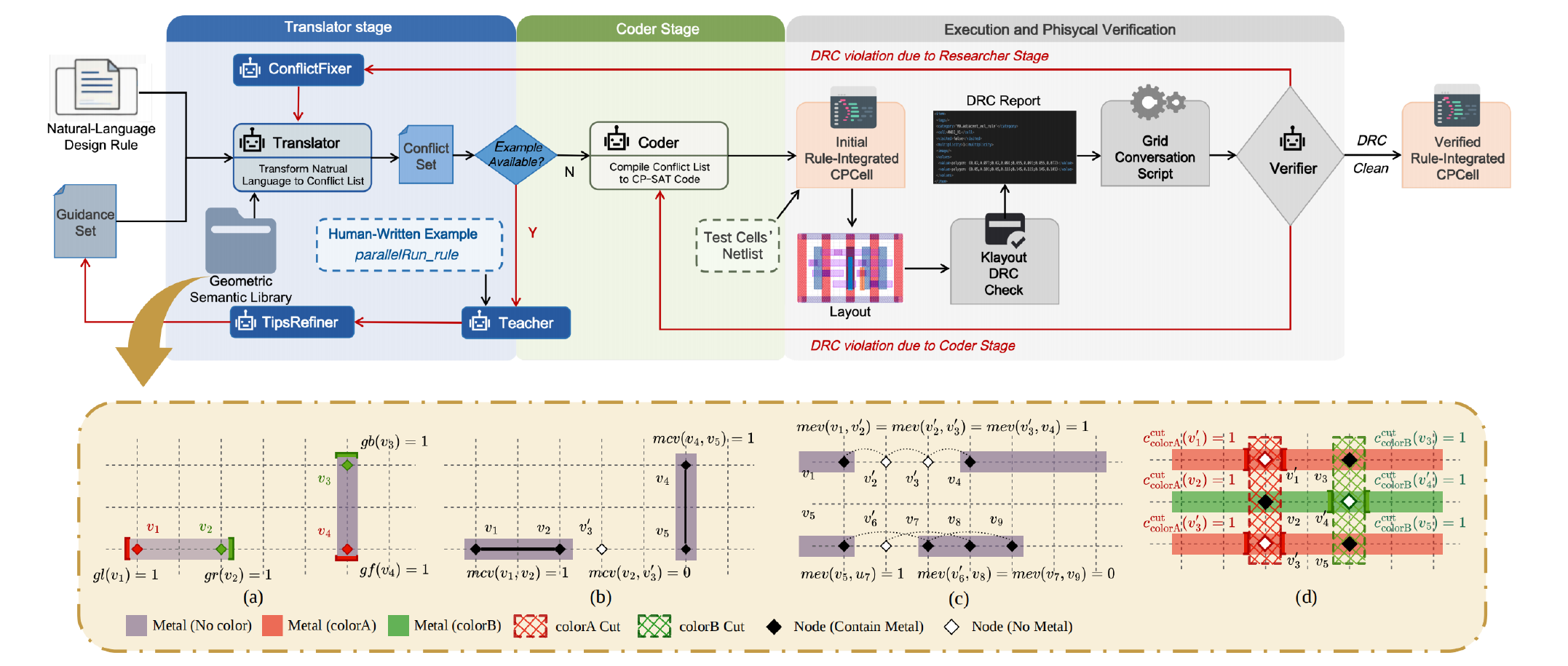}
  \caption{Overview of AutoDRI: the upper part shows the multi-agent automated integration flow, while the lower part illustrates the geometric semantic library over the routed grid.}
  \label{fig:overall_framework}
  \vspace{-0.5em}
\end{figure*}

\subsection{Design Rules and Representation Limitation}

Two distinct manufacturing mechanisms exist in commercial
PDKs.  
In a \emph{metal-based} encoding, the lithographic mask directly defines discrete metal segments, so rules are naturally expressed as spacing constraints between segment endpoints or between parallel metal shapes. 
In a \emph{cut-based} encoding, metal is first laid down as continuous tracks and then severed into individual segments by colored cut masks. 
Here, segment boundaries are created by cuts rather than by the metal mask itself. 
As a result, the governing rules shift from metal-endpoint spacing to cut-mask spacing: metal-based rules such as \texttt{eol\_rule} and \texttt{parallelRun\_rule} may become implicitly satisfied by the cut geometry, while new rules are introduced to constrain the cut masks directly. 
Under \emph{multi-patterning}, adjacent tracks are assigned alternating colors (e.g., colorA and colorB) for manufacturability. 
In a cut-based process, the cut masks inherit these colors, so cut-mask spacing rules are defined per color.

Table~\ref{tab:rule_descriptions} summarizes the design rules
considered in this work~\cite{chava_standard_2015,ma_sadp_2012} into four categories:
(1) end-of-line spacing rules (e.g., \texttt{eol\_rule}, \texttt{adjacent\_eol\_rule}); (2) via interaction rules (e.g., \texttt{via\_stack\_rule}, \texttt{via\_spacing\_rule}); (3) context-dependent metal length rules (e.g., \texttt{mar\_rule}, \texttt{parallelRun\_rule}, \texttt{step\_rule}); and (4) colored cut-mask spacing rules (e.g., \texttt{colorA\_cut\_spacing}, etc.). A single technology node may
mix both: the commercial sub-7nm setting in
Section~\ref{sec:experiment} applies cut-based rules on M0 while
retaining metal-based rules on M1 and M2.

Existing CO-based synthesis frameworks model geometry with a
single class of variable: a binary indicator that marks whether a
routing-grid node $u = (\ell,r,c)$ is a metal endpoint.
Categories~(1) and~(2) can be captured under this abstraction, but
categories~(3) and~(4) expose two representational gaps.
For category~(3), the representation is incomplete: context-dependent
metal-length rules require geometric relationships beyond endpoint
positions, and as illustrated in
Figure~\ref{fig:representation_limitation}(a), a
\texttt{parallelRun\_rule} violation whose overlap fully coincides
with one metal segment cannot be identified from endpoint indicators alone and
therefore goes unmodeled.
For category~(4), the limitation is more fundamental: colored
cut-mask spacing rules
(Figure~\ref{fig:representation_limitation}(b)) cannot be
represented at all, because existing formulations introduce no
variables for cut masks or cut-mask color assignments.
These gaps motivate the geometric semantic library and the
constructive multicolor-cut modeling introduced in
Section~\ref{sec:semantic_representation}, which supply the
interval-level and cut-level variables needed to express all four
rule categories within a unified CP-SAT formulation.

\section{Semantic Rule Representation}
\label{sec:semantic_representation}

To address the representational gaps identified in
Section~\ref{sec:prelim}, AutoDRI builds on a geometric semantic
library that lifts primitive edge variables into reusable
higher-level semantic variables over the routed grid, as illustrated
in Figure~\ref{fig:overall_framework}.
The library further includes a constructive multicolor-cut modeling
method that derives color-specific cut-mask variables from these
same semantic constructs.
Together, these close both the incomplete-coverage gap for
context-dependent metal-length rules and the missing-abstraction
gap for colored cut-mask spacing rules, while providing the
semantic basis for the downstream automated rule-integration flow.

\subsection{Geometric Semantic Library}

To obtain a more compact and expressive representation, we introduce a geometric semantic library over the routed 3D grid graph $G=(V,E)$, as illustrated in the lower part of Figure~\ref{fig:overall_framework}. 
Following the boundary-marker formulation adopted in SP\&R~\cite{lee_spr_2021}, we retain \texttt{geometric\_vars} as lightweight indicators of segment boundaries, as shown in Figure~\ref{fig:overall_framework}(a). 
For a node $u$ on a horizontal-routing layer, \texttt{gl}$(u)$ and \texttt{gr}$(u)$ indicate that $u$ is the left or right boundary of a metal segment, respectively; for a node $u$ on a vertical-routing layer,
$\texttt{gf}(u)$ and $\texttt{gb}(u)$ indicate the front and back boundaries. 
These variables identify where a segment starts or ends, while the underlying edge variables in $E$ still encode
primitive metal or via occupancy.

On top of these boundary markers, we introduce two interval-level semantic variables that are central to rule expression, as shown in Figure~\ref{fig:overall_framework}(b) and Figure~\ref{fig:overall_framework}(c). 
The first is the metal continuation variable \texttt{mcv}$(u,v)$, which denotes that two same-track nodes $u$ and $v$ belong to one continuous metal segment with no internal split. 
The second is the metal empty variable \texttt{mev}$(u,v)$, which denotes that the interval between $u$ and $v$ on the same track contains no metal.
Both are derived from the primitive edge variables along the same-track path $P(u,v)$, where $P(u,v)$ denotes the set of all edges on the same track between nodes $u$ and $v$:
\begin{equation}
\texttt{mcv}(u,v)=\bigwedge_{e\in P(u,v)} e,
\qquad
\texttt{mev}(u,v)=\bigwedge_{e\in P(u,v)} \neg e .
\end{equation}
To support cut-based multi-patterning process nodes, we further extend the library with a color-specific cut-mask variable $c^{\mathrm{cut}}_\kappa(u)$
for each color $\kappa$, indicating that the physical cut mask of
color~$\kappa$ passes through node~$u$
(see Figure~\ref{fig:overall_framework}(d)). Unlike the boundary and
interval variables above, $c^{\mathrm{cut}}_\kappa(u)$ cannot be
read off directly from primitive edge variables; its construction
requires an explicit modeling step, which we present in
Section~\ref{sec:cut_modeling}.  Together, the boundary markers,
interval variables, and cut-mask variables form the semantic basis
over which all design-rule constraints in this work are expressed.

\subsection{Constructive Multicolor-Cut Modeling}\label{sec:cut_modeling}

We now show how the color-specific cut-mask variables
$c^{\mathrm{cut}}_\kappa(u)$ are constructed.  The central
difficulty is that a physical cut mask may extend across nodes whose
intrinsic colors differ from the mask color, so the color of a cut mask
cannot be inferred from node colors alone.  We address this with a
two-step construction: first selecting color-independent cut
locations, then lifting them into color-specific cut-mask variables.

We begin by defining a node-color function.  For each layer~$\ell$,
let $O_\ell = (\kappa_0, \kappa_1, \ldots, \kappa_{m-1})$ denote its
cyclic color order (in practice, two colors \texttt{colorA} and
\texttt{colorB}).  For a node $u = (\ell, r, c)$, its intrinsic color
is assigned by track index:
\begin{equation}
\chi(u)=
\begin{cases}
O_\ell\!\left[\mathrm{idx}_{R_\ell}(r)\bmod |O_\ell|\right],
  & \text{for horizontal layer}~\ell,\\[0.3em]
O_\ell\!\left[\mathrm{idx}_{C_\ell}(c)\bmod |O_\ell|\right],
  & \text{for vertical layer}~\ell,
\end{cases}
\end{equation}
where $R_\ell$ and $C_\ell$ are the sorted row and column lists of
layer~$\ell$, and $\mathrm{idx}$ returns the zero-based position in
the corresponding list.

\begin{figure}[t]
  \centering

  \includegraphics[width=1.1\columnwidth]{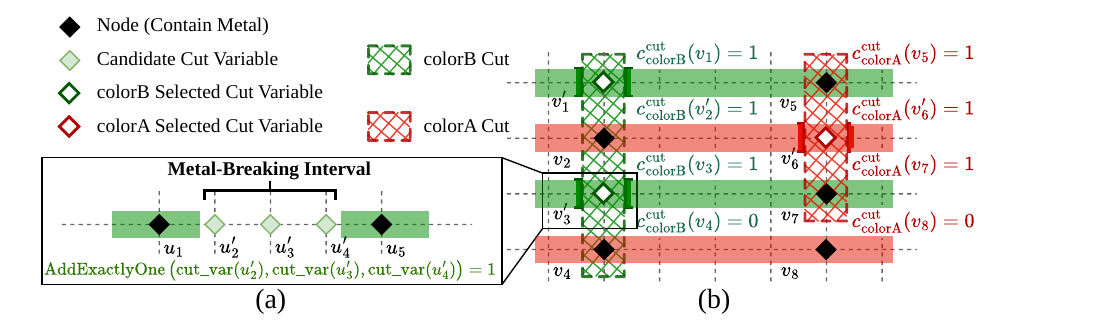}

  \caption{Constructive multicolor-cut modeling that derives color-specific cut-mask variables from primitive semantic variables.}
  \label{fig:multicolor_cut_model}
  \vspace{-0.8em}
\end{figure}

\textit{Step~1: Color-independent cut selection.}\quad
We introduce a variable $\texttt{cut\_var}(u)$ indicating that a cut
is selected at candidate node~$u$.  This variable marks where a
metal break is needed but does not yet carry color information.  A
valid $\texttt{cut\_var}$ must lie inside a \emph{metal-breaking interval}
--- an interval bracketed by two facing segment boundaries and
containing no metal
(Figure~\ref{fig:multicolor_cut_model}(a)).

On a horizontal layer, let $u_1 = (\ell,r,c_1)$ and
$u_5 = (\ell,r,c_5)$ with $c_1 < c_5$.  The candidate cut set
inside the interval is
\begin{equation}
\mathcal{Q}(u_1,u_5)
= \bigl\{(\ell,r,c) \;\bigm|\;
    c \in \mathrm{CutCols}_\ell,\; c_1 < c < c_5 \bigr\},
\end{equation}
where $\mathrm{CutCols}_\ell$ is the set of columns on layer~$\ell$
at which a cut may legally be placed.  An auxiliary Boolean variable
$h(u_1,u_5)$ indicates that $(u_1,u_5)$ forms a valid
metal-breaking interval:
\begin{equation}
h(u_1,u_5)
\;\Leftrightarrow\;
\texttt{mev}(u_1,u_5) \wedge \texttt{gr}(u_1) \wedge \texttt{gl}(u_5).
\end{equation}
If $\mathcal{Q}(u_1,u_5) \neq \varnothing$, exactly one cut must be
selected inside the interval:
\begin{equation}
h(u_1,u_5)
\;\Rightarrow\;
\sum_{w \in \mathcal{Q}(u_1,u_5)} \texttt{cut\_var}(w) = 1.
\end{equation}
Conversely, every activated $\texttt{cut\_var}$ must be justified by
at least one such interval:
\begin{equation}
\texttt{cut\_var}(w)
\;\Rightarrow\;
\bigvee_{(u_1,u_5):\, w \in \mathcal{Q}(u_1,u_5)} h(u_1,u_5).
\end{equation}
Together, $\texttt{mev}$ identifies where a legal metal gap exists,
while $\texttt{cut\_var}$ selects one concrete cut location within
that gap.

\textit{Step~2: Color lifting.}\quad
We now lift $\texttt{cut\_var}(u)$ into the color-specific variable
$c^{\mathrm{cut}}_\kappa(u)$.  If $\chi(u) = \kappa$, the node
already matches the mask color and the assignment is direct:
\begin{equation}
c^{\mathrm{cut}}_\kappa(u) = \texttt{cut\_var}(u),
\qquad \text{if } \chi(u) = \kappa.
\end{equation}
In Figure~\ref{fig:multicolor_cut_model}(b), activating
$\texttt{cut\_var}$ at nodes $v_1'$ and $v_3'$ directly induces
$c^{\mathrm{cut}}_B(v_1')$ and $c^{\mathrm{cut}}_B(v_3')$.

If $\chi(u) \neq \kappa$, the color-$\kappa$ cut mask may still
occupy node~$u$ provided it is supported by an activated
$\texttt{cut\_var}$ on a neighboring color-$\kappa$ node along the
same cut line.  Let $\mathcal{N}_\kappa(u)$ denote this set of
supporting neighbors; then
\begin{equation}
c^{\mathrm{cut}}_\kappa(u)
\;\Rightarrow\;
\bigvee_{v \in \mathcal{N}_\kappa(u)} \texttt{cut\_var}(v),
\qquad \text{if } \chi(u) \neq \kappa.
\end{equation}
For example, in Figure~\ref{fig:multicolor_cut_model}(b),
activating $\texttt{cut\_var}$ at node $v_6'$ supports
$c^{\mathrm{cut}}_A(v_5)$ and $c^{\mathrm{cut}}_A(v_7)$, even
though $v_5$ and $v_7$ do not intrinsically belong to color~$A$.
This mechanism allows a color-$\kappa$ cut mask to extend across
off-color nodes, anchored by a $\texttt{cut\_var}$ on a neighboring
color-$\kappa$ node.

With these variables in place, colored cut-mask spacing rules are
expressed directly as constraints over
$c^{\mathrm{cut}}_\kappa(u)$; the output writer merges contiguous
activated states into physical cut-mask geometries in the final
layout (e.g., $c^{\mathrm{cut}}_A(v_5) = c^{\mathrm{cut}}_A(v_6')
= c^{\mathrm{cut}}_A(v_7) = 1$ yields a single \texttt{colorA} mask from
$v_5$ to $v_7$).  The optimization objective simultaneously
minimizes the number of $\texttt{cut\_var}$ activations to suppress
unnecessary metal breaks and maximizes the active
$c^{\mathrm{cut}}_\kappa(u)$ variables to encourage each accepted
cut to extend into the longest legal physical mask segment.

\section{Automated Design Rule Integration}
\label{sec:automated_rule_integration}

Built on the geometric semantic library introduced in Section~\ref{sec:semantic_representation}, AutoDRI performs automated design rule integration through two key components: a
standardized conflict-set encoding that provides a uniform
representation of design rules over the semantic variables, and a multi-agent integration
flow that translates natural-language rules into executable CP-SAT
code.

\subsection{Conflict-Set Encoding}

Using the semantic variables in the geometric semantic library, we represent each design rule as a set of forbidden local Boolean patterns, which we call a \textit{conflict set}. 
For a design rule $\rho$, let
\begin{equation}
\Phi_{\rho}=\{\boldsymbol{\gamma}_1,\boldsymbol{\gamma}_2,\dots,\boldsymbol{\gamma}_M\}
\end{equation}
denote its conflict set, where each pattern $\boldsymbol{\gamma}_j=(b_1,b_2,\dots,b_t)$ is a tuple of Boolean literals describing one concrete violation scenario. 
A violation is triggered when all literals in a pattern are simultaneously true. 
The solver therefore forbids each such all-true local assignment by enforcing
\begin{equation}
\forall \boldsymbol{\gamma}_j \in \Phi_{\rho},\qquad
\sum_{b\in \boldsymbol{\gamma}_j} b \le |\boldsymbol{\gamma}_j|-1 .
\label{eq:conflict_list}
\end{equation}
Under this formulation, translating a design rule reduces to enumerating all forbidden local Boolean patterns that must never occur.

To translate a design rule into a conflict set, we use a standardized local-enumeration procedure: choose a reference node, enumerate all local violation scenarios around it, and emit one forbidden Boolean pattern for each scenario using the geometric semantic library.

\begin{figure}[htbp]
  \centering
  \includegraphics[width=\columnwidth]{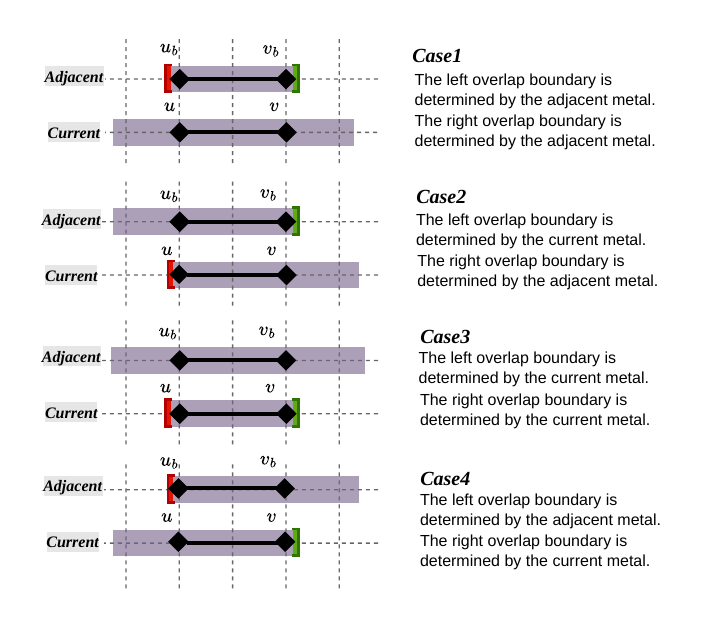}
  \caption{Four local violation scenarios of \texttt{parallelRun\_rule} on adjacent horizontal tracks.}
  \label{fig:conflict_list_encoding}
  \vspace{-0.6em}
\end{figure}

We use \texttt{parallelRun\_rule} on a horizontal routing layer as an example. 
Let $u$ be a reference node and consider all cases in which a too-short parallel overlap starts from $u$. 
As shown in Figure~\ref{fig:conflict_list_encoding}, this yields four local violation scenarios. 
All of them correspond to the same DRC condition, namely that the parallel overlap length is smaller than the required threshold, i.e., $\mathrm{dist}(u,v)<D$, where $D$ is the minimum legal parallel-run length.

These four cases can be expressed using only four semantic ingredients: a left overlap endpoint, a right overlap endpoint, and two continuous metal segments on adjacent tracks. 
Accordingly, they are encoded as the following four forbidden Boolean patterns, where $u$ and $v$ are two nodes whose distance is smaller than $D$, and $u_b$ and $v_b$ denote their respective back adjacent nodes:
\begin{equation}
\begin{aligned}
\boldsymbol{\gamma}_{\mathrm{PR}\mbox{-}1} &= \big(\texttt{gl}(u_b),\ \texttt{gr}(v_b),\ \texttt{mcv}(u_b,v_b),\ \texttt{mcv}(u,v)\big),\\
\boldsymbol{\gamma}_{\mathrm{PR}\mbox{-}2} &= \big(\texttt{gl}(u_b),\ \texttt{gr}(v),\ \texttt{mcv}(u_b,v_b),\ \texttt{mcv}(u,v)\big),\\
\boldsymbol{\gamma}_{\mathrm{PR}\mbox{-}3} &= \big(\texttt{gl}(u),\ \texttt{gr}(v),\ \texttt{mcv}(u_b,v_b),\ \texttt{mcv}(u,v)\big),\\
\boldsymbol{\gamma}_{\mathrm{PR}\mbox{-}4} &= \big(\texttt{gl}(u),\ \texttt{gr}(v_b),\ \texttt{mcv}(u_b,v_b),\ \texttt{mcv}(u,v)\big).
\end{aligned}
\label{eq:horizontal_parallel_run_conflicts}
\end{equation}

This example illustrates the simplification enabled by conflict-set encoding. 
Instead of writing handcrafted solver constraints over primitive variables, we derive the rule through local enumeration around a reference node and express each case as a standardized Boolean pattern over semantic variables. 
The result is a lightweight, modular, and uniform rule-translation process.

\subsection{Multi-Agent Integration Flow}
\begin{figure}[t]
  \centering
  \includegraphics[width=\columnwidth]{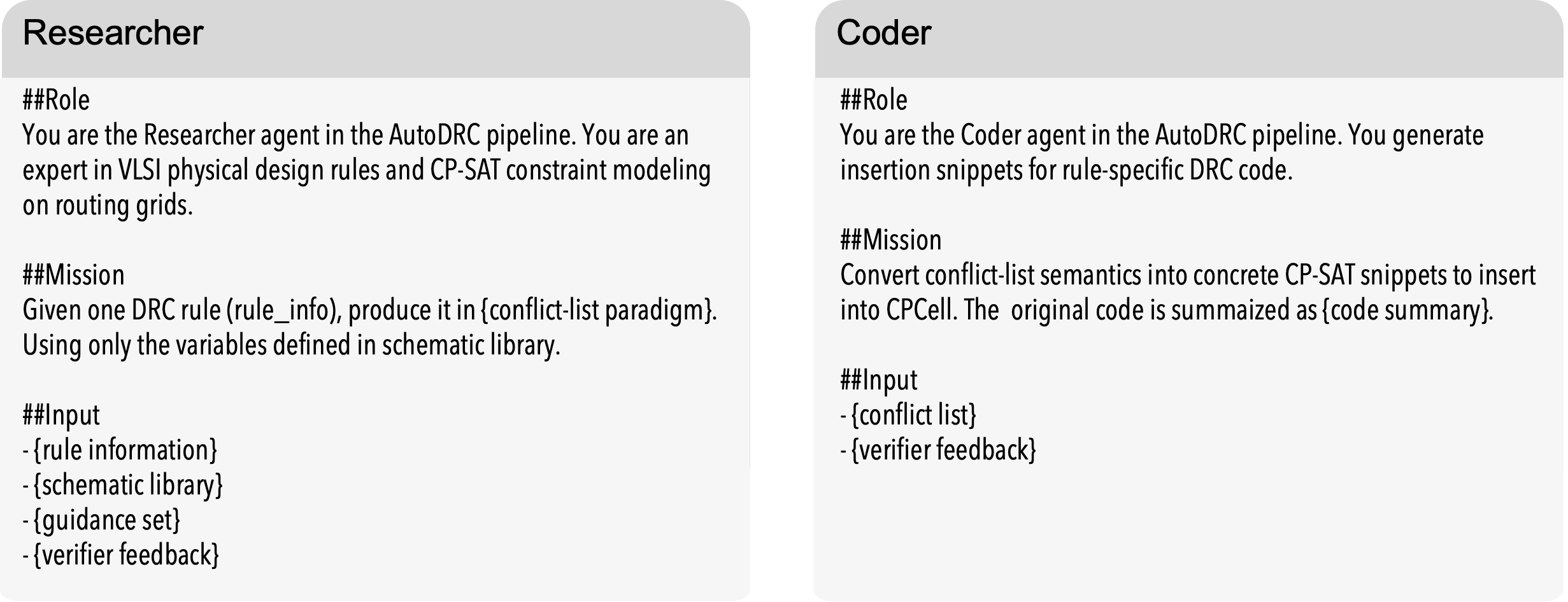}
  \caption{Core prompt structures of the two main agents.}
  \label{fig:core_agent_role}
  \vspace{-0.6em}
\end{figure}

In this section, we introduce our automated design rule integration flow with a multi-agent framework.
The upper part of Figure~\ref{fig:overall_framework} illustrates the overall flow. 
The flow consists of two consecutive translation stages from natural-language design rules to executable solver constraints. 
In the \textit{translator} stage, a natural-language design rule is translated into a standardized conflict-set representation. 
In the \textit{Coder} stage, the conflict set is compiled into executable CP-SAT code and integrated into the synthesis framework. 
The roles of the core agents in these two stages are further illustrated in Figure~\ref{fig:core_agent_role}.

To improve reliability, we introduce two feedback mechanisms over these two stages. 
A \textit{Teacher--Translator Loop} improves the translation from natural language to conflict sets, while a \textit{Verifier Feedback} mechanism analyzes synthesis and verification errors and feeds the resulting error information back to the responsible stage.

In the \textit{Teacher--Translator Loop}, the Translator generates a conflict-set representation $\Phi_{\rho}$ from the natural-language design rule under the current guidance set. 
For rules with human-written conflict-set demonstrations $\Phi_{\rho}^\ast$, the Teacher compares the generated conflict set with the reference one, and returns a match status, diagnostic feedback, and, when a semantic discrepancy is found, a set of concise natural-language tips $U^{(k)}$ for subsequent iterations. 
This follows the reflection-based refinement style used in prior iterative-improvement frameworks~\cite{shinn_reflexion_2023,madaan_self-refine_2023}. 
To convert these raw tips into persistent guidance, we further introduce a \textit{TipsRefiner}, inspired by SCOPE~\cite{pei_scope_2025}. 
Algorithm~\ref{alg:tipsrefiner} summarizes this update process.

Specifically, for each rule $r$, the framework maintains a tip memory $M_r$ that stores accumulated tips together with their feedback history across iterations. 
After each Teacher comparison, $M_r$ is updated using the newly generated tips $U^{(k)}$ and the current diagnostic feedback. 
A candidate tip pool $\mathcal{P}^{(k)}$ is then constructed from $M_r$ by re-evaluating recorded tips according to their empirical usefulness, such as whether the associated issue is resolved, persists, or repeatedly reappears in later iterations. 
Given $\mathcal{P}^{(k)}$, \textsc{Refine} updates the guidance set $G$ under a fixed prompt budget $B$ by filtering weak tips, merging semantically similar ones, and rewriting retained ones into compact guidance entries. 
The updated guidance set is used in the next Translator iteration. 

\begin{algorithm}[t]
\caption{Refining Teacher Tips into Guidance}
\label{alg:tipsrefiner}
\begin{algorithmic}[1]
\REQUIRE Rule $r$, current guidance set $G$, reference conflict set $\Phi_{\rho}^\ast$, tip memory $M_r$, budget $B$
\ENSURE Updated guidance set $G$

\FOR{iteration $k=1,2,\dots$}
    \STATE $\Phi_{\rho} \leftarrow \textsc{Translator}(r, G)$
    \STATE $(\texttt{status}, \texttt{feedback}, U^{(k)}) \leftarrow \textsc{TeacherCompare}(\Phi_{\rho}, \Phi_{\rho}^\ast)$
    \STATE $M_r \leftarrow \textsc{UpdateTipMemory}(M_r, U^{(k)}, \texttt{feedback})$
    \STATE $\mathcal{P}^{(k)} \leftarrow \textsc{ConstructCandidateTips}(M_r)$
    \STATE $G \leftarrow \textsc{Refine}(G, \mathcal{P}^{(k)}, B)$
    \IF{\texttt{status} = \texttt{match}}
        \STATE \textbf{break}
    \ENDIF
\ENDFOR
\end{algorithmic}
\end{algorithm}

In the \textit{Verifier Feedback} stage, the generated code is executed in the CPCell synthesis framework and evaluated through compilation, runtime behavior, and post-layout physical checks. 
The reported DRC violations are first converted by a script into the gridded representation used by CPCell synthesis, so that checker feedback can be aligned with the internal routing-grid model. 
The Verifier then determines whether the observed gridded violation pattern is already covered by the forbidden cases encoded in the current conflict set. 
If the observed violation is not prohibited by the existing conflict set, this indicates that the conflict-set generation is incorrect or incomplete. 
In this case, the gridded error information is routed back to the translator side, where a \textit{ConflictFixer} revises the original conflict set by adding or correcting the missing forbidden patterns. 
If the conflict set is semantically correct but the emitted code is incorrect or incomplete, the error information is passed directly to the Coder, which revises the existing implementation accordingly. 
This diagnostic-and-repair process again follows a reflection-style design, where execution feedback is treated as an informative signal for targeted revision rather than blind retry~\cite{shinn_reflexion_2023}.

Together, these two feedback mechanisms form a closed-loop rule-integration framework. 
The \textit{Teacher--Translator Loop} continuously refines semantic translation, while the \textit{Verifier Feedback} mechanism grounds revision in synthesis and verification evidence. 
This enables the framework to progressively converge toward correct and executable rule integration.
In our workflow, a single demonstrated rule, \texttt{parallelRun\_rule}, already provides useful guidance for the remaining benchmark rules.

\begin{table}[hbtp]
  \centering
  \caption{Evaluated Sub-7nm Design Rules}
  \label{tab:rule_descriptions}
  \resizebox{0.48\textwidth}{!}{
    \begin{tabular}{@{}lp{6cm}@{}}
      \toprule
      \textbf{Rule Name} & \textbf{Description} \\ 
      \midrule
      \texttt{parallelRun\_rule} & Minimum overlap distance for two metal segments on adjacent tracks. \\
      \texttt{eol\_rule} & Minimum routing direction distance between two facing endpoints on the same track. \\
      \texttt{mar\_rule} & Minimum length of a continuous metal segment on its track. \\
      \texttt{step\_rule} & Minimum routing direction distance between ends in the same direction on adjacent tracks. \\
      \texttt{adjacent\_eol\_rule} & Minimum routing direction distance between two facing endpoints on adjacent tracks. \\
      \texttt{via\_spacing\_rule} &  Minimum via-to-via Manhattan distance\\
      \texttt{via\_stack\_rule} &  Prohibition of overlapping projections for vias in vertically adjacent layers on their shared metal track\\
      \midrule
      \texttt{colorA\_cut\_spacing} & Perpendicular distance between two overlapping \texttt{colorA} cuts. \\
      \texttt{colorB\_cut\_spacing} & Perpendicular distance between two overlapping \texttt{colorB} cuts. \\
      \texttt{colorA\_cut\_to\_via\_spacing} & Distance between a \texttt{colorA} cut and an overlapping via on \texttt{colorA} metal. \\
      \texttt{colorB\_cut\_to\_via\_spacing} & Distance between a \texttt{colorB} cut and an overlapping via on \texttt{colorB} metal. \\
      \bottomrule
    \end{tabular}
  }
\end{table}

\section{Experimental Results}
\label{sec:experiment}

Our proposed multi-agent rule integration framework is implemented in Python and executed on a Linux workstation equipped with Intel Xeon Platinum 8260 CPUs. 
The underlying CP-SAT-based simultaneous placement and routing is solved using the Google OR-Tools CP-SAT solver~\cite{or_tools}. 
For physical verification, we utilize the open-source KLayout DRC engine and Cadance LVS to perform layout checks.

To evaluate generalizability and robustness, we use the 41 standard-cell masters in the PROBE benchmark suite~\cite{ma_sadp_2012,vashishtha_asap7pdk_2017,cheng_probe20_2022,kahng_probe_2018}, with cell sizes ranging from 2 to 28 transistors. 
We evaluate 11 complex design rules commonly encountered in sub-7nm nodes: the first seven are derived from the PROBE3 PDK, while the remaining four are abstracted from commercial 7nm specifications, as summarized in Table~\ref{tab:rule_descriptions}.

\begin{table*}[htbp]
  \centering
  \caption{Rule Translation Correctness Validation (Correct Runs out of 3 over 41 Cells).}
  \label{tab:correctness_validation}
  \resizebox{\textwidth}{!}{
  \begin{tabular}{@{}lcccccc@{}}
    \toprule
    \multirow{2}{*}{\textbf{Design Rule}} & \multicolumn{2}{c}{\textbf{Direct CP-SAT Translation (Baseline)}} & \multicolumn{2}{c}{\textbf{Semantic-Driven One-Shot (Ablation)}} & \multicolumn{2}{c}{\textbf{Semantic-Driven Multi-Agent (Proposed)}} \\
    \cmidrule(lr){2-3} \cmidrule(lr){4-5} \cmidrule(lr){6-7}
    & \textbf{GPT-5.4} & \textbf{Gemini-3-pro} & \textbf{GPT-5.4} & \textbf{Gemini-3-pro} & \textbf{GPT-5.4} & \textbf{Gemini-3-pro} \\
    \midrule
    \texttt{eol\_rule}            & 2/3 & 2/3 & 3/3 & 3/3 & 3/3 & 3/3 \\
    \texttt{mar\_rule}            & 2/3 & 3/3 & 3/3 & 3/3 & 3/3 & 3/3 \\
    \texttt{parallelRun\_rule}    & 1/3 & 1/3 & 3/3 & 2/3 & 2/3 & 3/3 \\
    \texttt{step\_rule}           & 0/3 & 2/3 & 1/3 & 2/3 & 3/3 & 3/3 \\
    \texttt{adjacent\_eol\_rule}  & 1/3 & 2/3 & 2/3 & 0/3 & 3/3 & 3/3 \\
    \texttt{via\_spacing\_rule}   & 3/3 & 3/3 & 3/3 & 3/3 & 3/3 & 3/3 \\
    \texttt{via\_stack\_rule}     & 1/3 & 1/3 & 3/3 & 3/3 & 3/3 & 3/3 \\
    \texttt{colorA\_cut\_spacing} & 0/3 & 0/3 & 2/3 & 3/3 & 3/3 & 3/3 \\
    \texttt{colorB\_cut\_spacing} & 0/3 & 0/3 & 3/3 & 3/3 & 3/3 & 3/3 \\
    \texttt{colorA\_cut\_to\_via\_spacing} & 0/3 & 0/3 & 2/3 & 1/3 & 3/3 & 3/3 \\
    \texttt{colorB\_cut\_to\_via\_spacing} & 0/3 & 0/3 & 2/3 & 2/3 & 3/3 & 3/3 \\
    \midrule
    \textbf{Rules with 3/3 Correct Runs} & \textbf{1 / 11} & \textbf{2 / 11} & \textbf{6 / 11} & \textbf{6 / 11} & \textbf{10 / 11} & \textbf{11 / 11} \\
    \textbf{Correct Integrations over 33 Runs} & \textbf{10 / 33} & \textbf{14 / 33} & \textbf{27 / 33} & \textbf{25 / 33} & \textbf{32 / 33} & \textbf{33 / 33} \\
    \bottomrule
  \end{tabular}
  }
\end{table*}

\subsection{Overall Correctness Validation}
To evaluate our methodology and isolate the contributions of semantic abstractions versus the multi-agent feedback mechanism, we compare three configurations:

\begin{itemize}[noitemsep, topsep=0pt, leftmargin=*]
    \item \textbf{Direct CP-SAT Translation (Baseline):} direct translation from natural-language design rules to primitive solver variables using non-standardized \textit{at-most-one} constraints, similar to SP\&R~\cite{lee_spr_2021}.
    \item \textbf{Semantic-Driven One-Shot (Ablation):} a two-stage strategy that utilizes our geometric semantic library and conflict-set encoding but performs translation in a one-shot manner.
    \item \textbf{Semantic-Driven Multi-Agent (Proposed):} our full framework leveraging agent collaboration and the multi-stage feedback loop.
\end{itemize}

We evaluate all three configurations using Gemini-3-pro and GPT-5.4, with three runs per configuration to reduce stochastic variation. 
For each target design rule, AutoDRI generates a rule-integrated CP-SAT-based cell synthesis engine and uses it to synthesize all 41 standard-cell test cases. 
The resulting layouts then undergo design rule checking and physical verification. 
A run is considered correct only if the generated engine compiles successfully, all 41 synthesized cells are free of violations of the target rule, and all layouts pass LVS. 
Table~\ref{tab:correctness_validation} reports, for each rule, the number of correct runs out of three.

The experimental results demonstrate the superiority of our proposed methodology. Key analytical observations include:

{
\interlinepenalty=0
\predisplaypenalty=0
\postdisplaypenalty=0
\brokenpenalty=0

\begin{itemize}[noitemsep, topsep=0pt, leftmargin=*]

    \item \textbf{Normalization via Semantic Abstractions:} Comparing the Baseline with the Ablation demonstrates that our geometric semantic library and conflict-set encoding provide a structured bridge from high-level rule semantics to executable constraint patterns, reducing hallucination and lowering the difficulty of expressing complex rules.    
    
    \item \textbf{Stability via Multi-Agent Feedback:} Comparing the Ablation with the Proposed configuration shows that the \textit{Teacher--translator} loop and the \textit{Verifier Feedback} mechanism are essential for stabilizing the flow. By synergizing these strategies, the framework pushes ``mostly correct'' initial logic to ``fully correct'' rule integration across our benchmark set in the reported experiments.
    
    \item \textbf{Robustness across Reasoning Engines:} Under the full Multi-Agent framework, both GPT-5.4 and Gemini-3-pro achieve consistently high correctness in the reported experiments, reaching 32/33 and 33/33 correct integrations, respectively. This indicates that our framework remains robust across different reasoning engines and supports reliable design rule integration.
    
\end{itemize}
}

\subsection{Case Study: Process Migration Across Technology Nodes}
To evaluate the process-migration capability of AutoDRI, we conduct a case study on two distinct rule sets. 
\textbf{Rule List 1} follows an ASAP7-style metal-based encoding~\cite{xu_standard_2017,ASAP7}, where design rules are enforced through direct metal spacing. 
\textbf{Rule List 2} follows a commercial sub-7nm-style cut-based encoding, where the M0 layer is governed by colored cut-mask spacing rules, while M1 and M2 retain the metal-spacing rules from Rule List 1. 
The specific parameters of the two encodings are summarized in Table~\ref{tab:case_study_rules}. 
In Rule List 2, M0 rules such as \texttt{eol\_rule} and \texttt{parallelRun\_rule} are omitted because they are implicitly enforced through the constructive cut-mask constraints.

\begin{figure}[t]
  \centering

  \includegraphics[width=1.06\columnwidth]{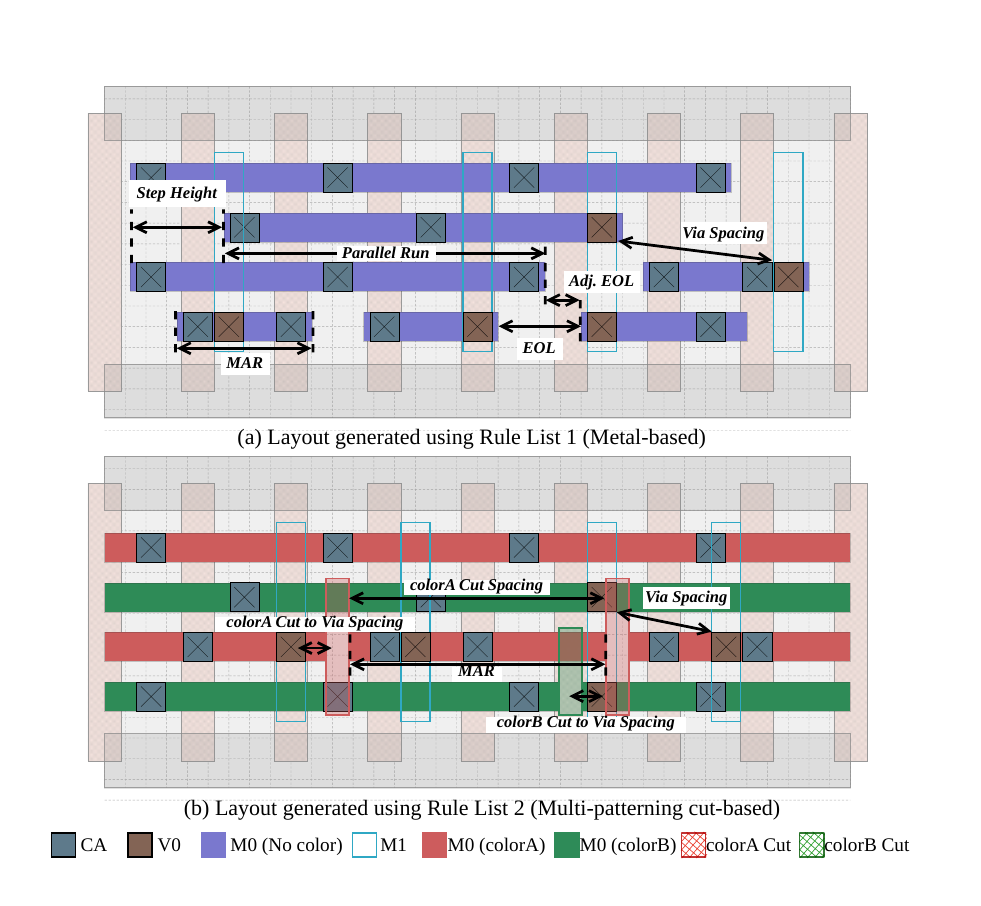}

  \caption{GDSII layout comparison of the \texttt{AOI21\_X2} standard cell across two distinct encodings. By modularly extending the geometric semantic library, AutoDRI successfully automates rule integration for disparate technology nodes.}
  \label{fig:case_study_aoi22}
\end{figure}

To support Rule List 2, we only need to extend the geometric semantic library with cut-aware and color-aware variables through the proposed constructive multicolor-cut modeling method. 
The core AutoDRI flow and the underlying solver formulation remain unchanged. 
With this extension, both rule sets can be automatically translated into executable CP-SAT constraints and integrated into the CPCell synthesis framework. 
The generated layouts for all 41 cell benchmarks pass KLayout DRC and Cadence LVS under both encodings.

Figure~\ref{fig:case_study_aoi22} shows the synthesized \texttt{AOI21\_X2} layouts under the two encodings. 
To highlight the M0-layer differences, components such as M1--M2 vias, M2 metal segments, and MEOL-level layouts are omitted. 
The physical realizations are substantially different: Rule List 1 yields segmented metal tracks to satisfy metal-spacing constraints, whereas Rule List 2 yields continuous colored metal tracks that are later severed by corresponding colored cut masks. 
With a grid resolution of 10\,nm, the geometric spacings shown in Figure~\ref{fig:case_study_aoi22} can be directly compared with the rule parameters in Table~\ref{tab:case_study_rules}, confirming that the illustrated layouts are DRC-clean under their respective encodings. In Table~\ref{tab:case_study_rules}, entries written as a single layer name denote same-layer metal rules, while the notation \texttt{via(M0,M1)} and \texttt{via(M1,M2)} denotes via-related rules.

This case study shows that AutoDRI can adapt the same CP-SAT-based synthesis framework to fundamentally different manufacturing encodings through a modular extension of the geometric semantic library, without rewriting the core integration flow.

\begin{table}[t]
  \centering
  \caption{Design Rule Parameters for Distinct Encodings (units: nm)}
  \label{tab:case_study_rules}
  \tblfont
  \setlength{\tabcolsep}{2.0pt}
  \renewcommand{\arraystretch}{1.0}
  \begin{tabular*}{\columnwidth}{@{\extracolsep{\fill}}llccc@{}}
    \toprule
    \textbf{Encoding} & \textbf{Rule Name} & \makecell[c]{\textbf{M0 /}\\\textbf{via(M0,M1)}} & \makecell[c]{\textbf{M1 /}\\\textbf{via(M1,M2)}} & \textbf{M2} \\
    \midrule
    \textbf{Rule List 1} & \texttt{mar\_rule} & 35 & 96 & 84 \\
    (Metal-based) & \texttt{parallelRun\_rule} & 50 & 48 & 54 \\
                   & \texttt{eol\_rule} & 10 & 12 & 6 \\
                   & \texttt{step\_rule} & 22 & 22 & 22 \\
                   & \texttt{adjacent\_eol\_rule} & 2 & 12 & 6 \\
                   & \texttt{via\_spacing\_rule} & 8 & 8 & - \\
    \midrule
    \textbf{Rule List 2} & \texttt{mar\_rule} & 35 & 96 & 84 \\
    (Cut-based) & \texttt{colorA\_cut\_spacing} & 37 & - & - \\
                & \texttt{colorB\_cut\_spacing} & 37 & - & - \\
                & \texttt{colorA\_cut\_to\_via\_spacing} & 4 & - & - \\
                & \texttt{colorB\_cut\_to\_via\_spacing} & 4 & - & - \\
                & \texttt{via\_spacing\_rule} & 8 & 8 & - \\
                & \texttt{eol\_rule} & - & 12 & 6 \\
                & \texttt{step\_rule} & - & 22 & 22 \\
                & \texttt{adjacent\_eol\_rule} & - & 12 & 6 \\
                & \texttt{parallelRun\_rule} & - & 48 & 54 \\
    \bottomrule
  \end{tabular*}
\end{table}

\begin{table}[t]
  \centering
  \caption{Runtime Comparison for Complex Cells}
  \label{tab:runtime_comparison}
  \tblfont
  \setlength{\tabcolsep}{2.6pt}
  \renewcommand{\arraystretch}{1.0}
  \begin{tabular*}{\columnwidth}{@{\extracolsep{\fill}}lcccc@{}}
    \toprule
    \textbf{Cell Name} & \textbf{\# FETs} & \textbf{Baseline (s)} & \textbf{Proposed (s)} & \textbf{Ratio} \\
    \midrule
    AND3\_X2   & 8  & 20.79   & 21.49   & $1.03\times$ \\
    AOI21\_X2  & 6  & 65.60   & 78.92   & $1.20\times$ \\
    AOI22\_X2  & 8  & 245.52  & 184.68  & $0.75\times$ \\
    BUF\_X8    & 4  & 309.39  & 269.82  & $0.87\times$ \\
    DFFHQN\_X1 & 24 & 900.52  & 1636.39 & $1.82\times$ \\
    DFFRNQ\_X1 & 28 & 8781.61 & 5346.98 & $0.61\times$ \\
    INV\_X8    & 2  & 61.82   & 56.08   & $0.91\times$ \\
    NAND3\_X2  & 6  & 119.70  & 105.30  & $0.88\times$ \\
    NAND4\_X2  & 8  & 213.53  & 157.65  & $0.74\times$ \\
    NOR3\_X2   & 6  & 91.78   & 69.26   & $0.75\times$ \\
    NOR4\_X2   & 8  & 156.93  & 151.74  & $0.97\times$ \\
    OAI21\_X2  & 6  & 85.78   & 77.40   & $0.90\times$ \\
    OAI22\_X2  & 8  & 164.94  & 181.84  & $1.10\times$ \\
    XOR2\_X1   & 10 & 22.33   & 20.98   & $0.94\times$ \\
    \midrule
    \textbf{Average} & \textbf{-} & \textbf{-} & \textbf{-} & \textbf{$0.96\times$} \\
    \bottomrule
  \end{tabular*}
\end{table}

\begin{table}[t]
  \centering
  \caption{Block-level Implementation Metrics}
  \label{tab:block-level}
  \tblfont
  \setlength{\tabcolsep}{1.2pt}
  \renewcommand{\arraystretch}{1.0}
  \begin{tabular*}{\columnwidth}{@{\extracolsep{\fill}}l c cc cc cc cc@{}}
    \toprule
    \multirow{3}{*}{\textbf{Design}} &
    \multirow{3}{*}{\textbf{\#Cell}} &
    \multicolumn{8}{c}{\textbf{Block-level Metrics}} \\
    \cmidrule(lr){3-10}
    & &
    \multicolumn{2}{c}{\makecell[c]{\textbf{Core Area}\\\textbf{($\mu$m$^2$)}}} &
    \multicolumn{2}{c}{\makecell[c]{\textbf{Total Power}\\\textbf{(mW)}}} &
    \multicolumn{2}{c}{\makecell[c]{\textbf{Eff. Clock}\\\textbf{(ns)}}} &
    \multicolumn{2}{c}{\textbf{DRVs}} \\
    \cmidrule(lr){3-4} \cmidrule(lr){5-6} \cmidrule(lr){7-8} \cmidrule(lr){9-10}
    & &
    \textbf{Base.} & \textbf{Ours} &
    \textbf{Base.} & \textbf{Ours} &
    \textbf{Base.} & \textbf{Ours} &
    \textbf{Base.} & \textbf{Ours} \\
    \midrule
    \texttt{gcd}           & 246   & 12.131   & 12.131   & 0.0295  & 0.0281  & 0.160 & 0.160 & 928    & 0 \\
    \texttt{aes}           & 10664 & 412.983  & 412.983  & 1.0200  & 1.0400  & 0.230 & 0.230 & 27554  & 0 \\
    \texttt{jpeg\_encoder} & 52557 & 2340.576 & 2331.439 & 10.4000 & 10.5000 & 0.230 & 0.220 & 168410 & 0 \\
    \bottomrule
  \end{tabular*}
\end{table}

\subsection{Runtime and Efficiency Analysis}

Although the semantic layer improves LLM comprehension, it may also increase solver size. 
We therefore compare runtime on 14 complex cells using the four rules already integrated in the original CPCell framework: \texttt{eol\_rule}, \texttt{mar\_rule}, \texttt{via\_spacing\_rule}, and \texttt{parallelRun\_rule}.
The comparison is performed between the following two formulations:
\begin{itemize}[noitemsep, topsep=0pt, leftmargin=*]
    \item \textbf{Baseline (Hard-coded At-Most-One):} the original CPCell formulation, which serves as the underlying framework for our rule-integration flow and manually encodes these four rules directly on primitive variables using \textit{at-most-one} constraints. 
    \item \textbf{Proposed (Semantic conflict-set encoding):} our automatically integrated formulation for the same four rules, generated using intermediate semantic variables and deduplicated conflict sets.
\end{itemize}

As summarized in Table~\ref{tab:runtime_comparison}, the proposed formulation achieves runtime that is overall comparable to the baseline, with an average ratio of $0.96\times$ across the 14 complex cells. 
We attribute this efficiency to the logical presolve capability of the CP-SAT engine. 
Because the conflict sets are strictly deduplicated and the helper variables maintain clear linear relationships with the underlying grid edges, the presolver can effectively collapse much of the semantic layer before branching begins.
These results indicate that the proposed formulation preserves competitive runtime in practice while providing the key benefit of automated, LLM-driven rule integration.

\begin{figure}[hbtp]
  \centering

  \includegraphics[width=0.95\columnwidth]{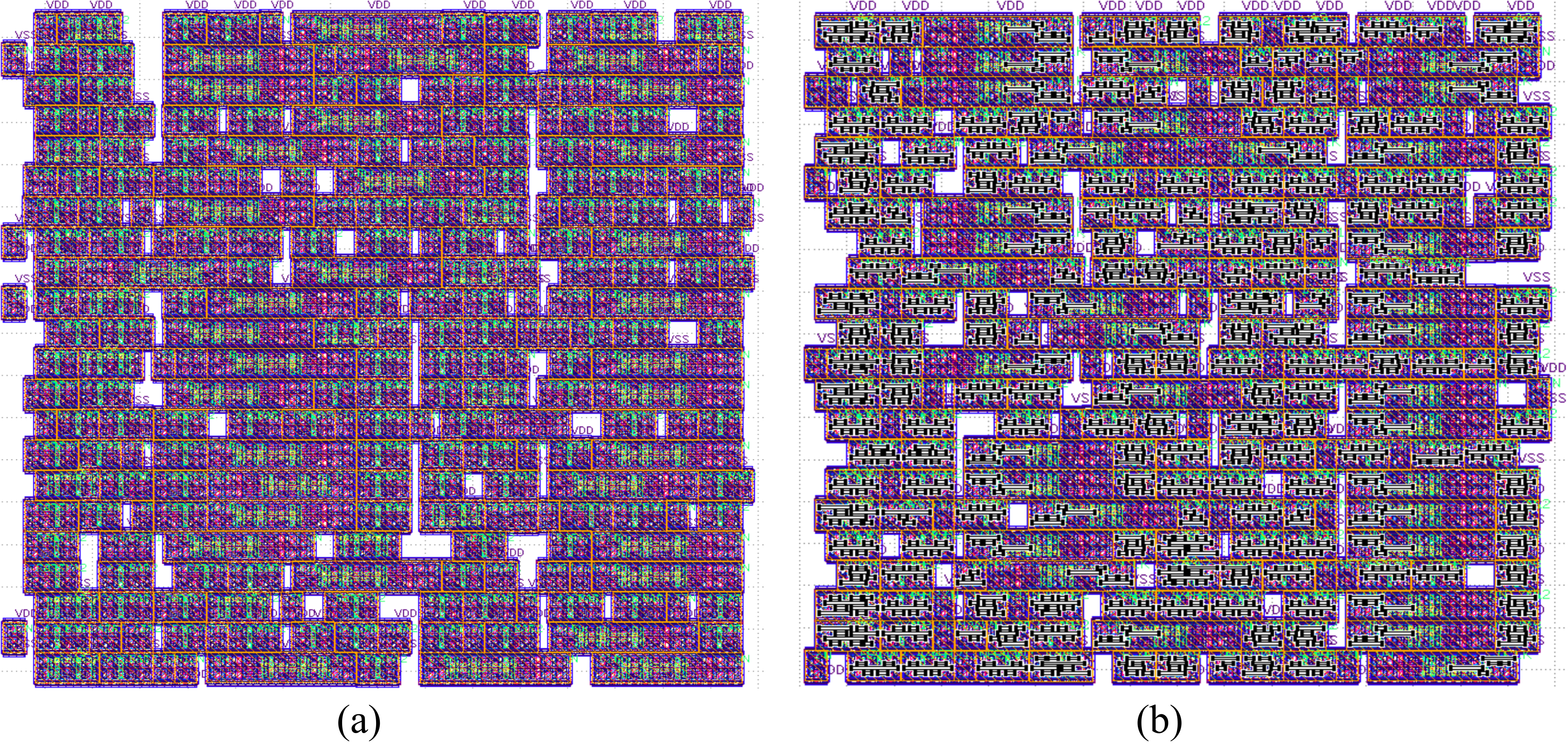}

  \caption{Block-level DRC marker comparison on the \texttt{gcd} design: (a) \textbf{Ours} and (b) \textbf{Baseline}.}
  \label{fig:drc_comparation}
\end{figure}

\subsection{Block-Level Validation}

To further evaluate the practical impact of our framework beyond cell-level correctness, we conduct a block-level validation using Rule List 1. 
We first automatically integrate the rules in Rule List 1 into the CPCell framework and synthesize the corresponding standard cells. 
Following a flow similar to PROBE3~\cite{choi_probe30_2024}, we then convert the synthesized cells into a usable PDK under the same virtual technology setting used in PROBE3.

At the block level, we compare two PDKs: \textbf{Baseline}, generated from the unmodified CPCell~\cite{CPCellUCSD} flow, and \textbf{Ours}, generated from our automated rule-integration flow. 
The comparison is performed on three representative designs, namely \texttt{gcd}, \texttt{aes}, and \texttt{jpeg\_encoder}. 
The resulting block-level implementation PPA metrics are summarized in Table~\ref{tab:block-level}. 
Here, the reported DRVs count only design-rule violations caused by standard-cell metal. 
The results show that the rules integrated by AutoDRI remain effective at the block level: cells generated by our flow maintain PPA comparable to the original CPCell-generated cells while eliminating the cell-metal DRVs observed in the baseline flow.

Figure~\ref{fig:drc_comparation} further compares the DRC markers of \textbf{Ours} in (a) and \textbf{Baseline} in (b), using the \texttt{gcd} design as an example. 
Here, the exported GDS contains only standard-cell metal shapes, and the black rectangles denote the resulting DRC markers caused by cell metal. 
The baseline flow exhibits many such violations because complex rules such as \texttt{parallelRun\_rule} are not integrated, whereas our flow automatically integrates these rules and therefore eliminates the cell-metal DRVs in this experiment.

\section{Conclusions}
\label{sec:conclusions}

This paper presents \textbf{AutoDRI}, a framework for automated design rule integration in CP-SAT-based standard cell synthesis. 
AutoDRI combines a geometric semantic library, a standardized conflict-set encoding, a constructive multicolor-cut modeling method, and a multi-agent integration flow to translate complex design rules into executable CP-SAT constraints within the CPCell framework.

There are two main directions for future work. 
First, we will extend AutoDRI to multimodal rule understanding by incorporating figures and diagrams from foundry documentation. 
Second, we will further develop AutoDRI from a rule-integration framework into a foundational module for larger DTCO/STCO workflows.

\end{document}